# Unveiling Nanoscale Surface Damage Dynamics in Swift Heavy Ion Irradiated Gallium Nitride

*Jiayu Liang[1], Shaowei He[1], Wenlong Liao[2,3], Tan Shi[1], Hang Zang[1], Yonghong Li[1], Wenbo Liu[1], Xiaojun Fu[2,3], Chuanjian Yao[2,3], Huan He[1,4*], Jianan Wei[2,3*] and Chaohui He[1,4*]*

[1] School of Nuclear Science and Technology, Xi'an Jiaotong University, Xi'an, 710049, China

[2] Academy of Chips Technology, China Electronics Technology Group Corporation, Chongqing, 400060, China

[3] National Key Laboratory of Integrated Circuits and Microsystems, Chongqing, 400060, China

[4] National Key Laboratory for High Energy Pulsed Power, Xi'an, 710049, China

[*]E-mail: huanhe@xjtu.edu.cn,weijianan93@163.com & hechaohui@xjtu.edu.cn

Abstract：

This work systematically unveils the nanoscale surface damage dynamics in gallium nitride by investigating the atomistic mechanisms of hillock formation. The results identify two distinct hillock morphologies dependent on electronic energy loss ($S_e$) values. Bell-shaped hillocks form under 18.2 keV/nm Kr irradiation, whereas crater-rim hillocks with central holes emerge under 40.2 keV/nm Ta irradiation. Microstructural analysis reveals that Ga-rich hillocks are accompanied by the generation of metastable zincblende nanodomains. These nanodomains preferentially aggregate around the periphery or sidewalls of the hillocks and exhibit a high spatial correlation with screw dislocations. Further temperature-dependent studies indicate that elevated temperatures significantly enlarge the overall dimensions of the hillock structures without altering their fundamental morphologies. Notably, under Ta irradiation above 1200 K, the high temperatures drastically reduce the viscosity and surface tension of liquid gallium. This enhanced fluidity of the transient molten phase promotes the formation of penetrating nanochannels.

Gallium nitride (GaN) has emerged as a core material for high-power, high-frequency, and radiation-resistant electronic devices owing to its wide bandgap, high breakdown field strength, and excellent thermal stability.[1–5] However, in aerospace and nuclear applications, GaN-based devices are inevitably exposed to swift heavy ions (SHIs) irradiation.[6,7] These high-energy particles deposit massive energy along their trajectories, generating intense thermal spikes. Subsequent melt-recrystallization processes lead to the formation of ion tracks, dislocations, and nanoscale defects, as well as pronounced surface modifications such as protrusions and craters.[8–10]

Advanced characterization techniques, such as scanning electron microscopy (SEM) and atomic force microscopy (AFM), have observed nanoscale hillocks and crater-like structures on material surfaces under SHI irradiation.[11–13] For GaN, Kumar *et al.* reported that surface features are significantly affected by nitrogen loss and damage overlaps under high-fluence 100 MeV Au ion irradiation.[14] Zhang *et al.* further demonstrated that the hillock diameter and density on GaN surfaces increase markedly with both ion fluence and surface roughness.[15] Beyond experiments, Sequeira *et al.* have substantially advanced the atomistic understanding of SHI-induced damage mechanisms in GaN through coupled two-temperature model and molecular dynamics (TTM-MD) simulations. They identified strong recrystallization as a core mechanism underlying the resistance of GaN to intense ionizing radiation and established an empirical relationship between the electronic energy loss ($S_e$) values of SHIs and the molten/amorphous cross-section.[16,17]

Nevertheless, current irradiation damage simulations predominantly focus on bulk phenomena, leaving the atomistic mechanisms governing the formation of irradiation-induced surface hillocks in GaN largely underexplored, where thermal gradients, atomic sputtering, and transient high-pressure effects play critical roles. Moreover, the highly dynamic processes of surface hillock formation and structural damage are inherently difficult to characterize clearly through direct experimental observations. Crucially, these fundamental damage dynamics are highly sensitive to thermal environment. In practical applications, GaN devices operating under high bias voltages often experience severe Joule heating, which creates transient extreme-temperature environments.[18–21] The coupling of SHI impacts with such heated environments profoundly accelerates material degradation and potentially catastrophic device failure. However, a clear microscopic physical picture detailing both the intrinsic dynamic formation of these surface hillocks and their subsequent evolution under synergistic extreme temperatures is still lacking. This fundamental knowledge gap severely restricts the widespread deployment of GaN devices in aerospace environments.

To investigate the atomistic formation mechanism of SHI-induced hillocks and their evolution under the synergistic effect of irradiation and transient high temperatures, we employed a multiscale simulation approach combining the two-temperature model (TTM) and molecular dynamics (MD). In this work, two typical SHIs, 430 MeV Kr and 1171 MeV Ta ions with corresponding $S_e$ values of 18.2 and 40.2 keV/nm, were utilized to simulate SHI irradiation. Irradiation was applied centrally along the [0001] crystallographic direction within a 30 × 30 × 100 nm$^3$ supercell. To capture the thermal transient effects, ambient temperatures were systematically varied across 300 K, 600 K, 900 K, 1200 K, 1500 K, and 1800 K. Additional simulation details are provided in Section 1 of the supplementary materials.

Figures 1(a) and (b) show schematic diagrams of ion tracks and atomic displacement in GaN under SHI irradiation with different $S_e$ values. The radius of the ion track gradually decreases with increasing depth, and the region near the surface and the ion incidence core exhibits more severe atomic displacement. Under 430 MeV Kr irradiation, scattered bubbles form exclusively in the central region extending from the surface down to a depth of approximately 18.4 nm. In contrast, under 1171 MeV Ta irradiation, an ion track consisting of a continuous nanochannel is generated roughly 28.7 nm in depth. As the ion penetration depth increases, this continuous nanochannel evolves into discontinuous nanoscale cavities with progressively diminishing sizes.

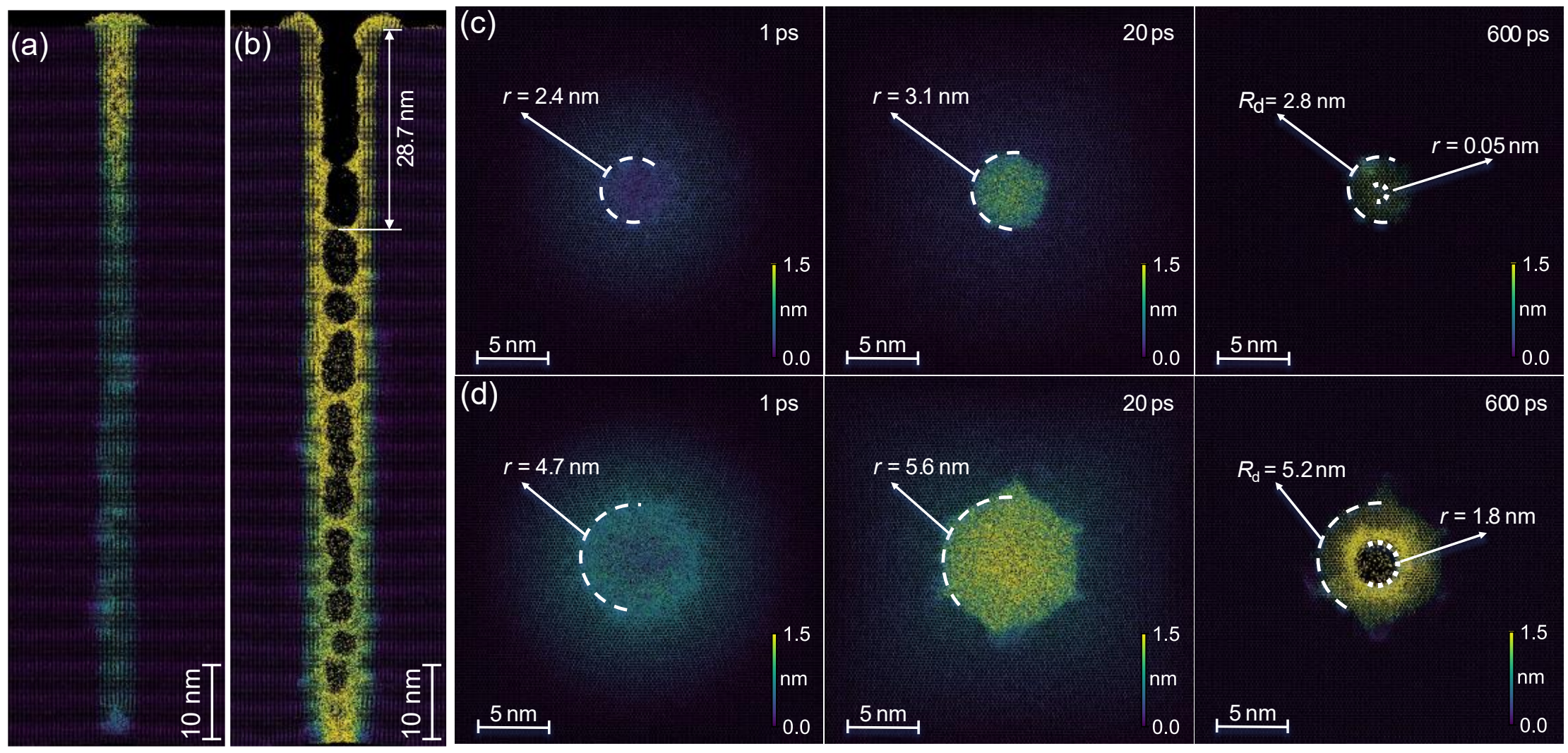

**FIG. 1.** Schematic diagrams of ion tracks in GaN irradiated by (a) 430 MeV Kr and (b) 1171 MeV Ta ions. Cross-sectional schematics (30 nm in thickness) at 700 nm depth along the ion incidence path under (c) 430 MeV Kr and (d) 1171 MeV Ta irradiation.

Here, $r$ denotes the radius of the amorphous molten region, and $R_d$ denotes the atomic displacement radius. The color scale represents the atomic displacement magnitude relative to the original lattice positions.

To characterize the melt-recrystallization process, the ion track evolution was analyzed on a 30-nm-thick cross-section at a sufficient depth of 700 nm from the surface. Cross-sectional schematics at three key stages following 430 MeV Kr and 1171 MeV Ta irradiation are presented in Figures 1(c) and (d), respectively. These snapshots reveal that at the initial 1 ps, the intense thermal spike induced by SHI incidence melts GaN lattice instantaneously, forming a cylindrical amorphous region centered along the ion trajectory. Within the first 20 ps, the ultrahigh temperature at the thermal spike center drives the radial expansion of the amorphous region until it reaches its maximum radius. In the subsequent evolution, the recrystallization process becomes dominant. Driven by thermal relaxation, the GaN lattice dissipates heat to the external environment, causing the overall temperature to decrease gradually. Recrystallization proceeds inward from the outer edge of the amorphous region toward ion track core. Once the system temperature drops to room temperature, recrystallization is essentially completed, ultimately leaving behind an ion track with distinct morphological characteristics. The evolution of wurtzite GaN density and the fractions of Ga hexagonal close-packed structure further corroborates this melt-recrystallization mechanism under SHI irradiation. Detailed analyses are available in our previous work[22] and Section 2 of the supplementary materials. At the 700 nm depth, the final ion track radius is approximately 0.05 nm for 430 MeV Kr irradiation and 1.8 nm for 1171 MeV Ta irradiation. These results confirm a positive correlation between the $S_e$ value of the incident SHI and the final radius of the amorphous region.

The dynamic formations of surface hillocks under 430 MeV Kr and 1171 MeV Ta irradiation are exhibited in Figures 2(a) and (b). Under 430 MeV Kr irradiation, the moderate thermal pressure induces only a weak surface perturbation at 1 ps, with a slight central uplift and minor sputtering. As the amorphous molten region expands within the first 20 ps, outward-migrating atoms form an upward-protruding incipient hillock. The height and width of the hillock increase simultaneously with radial expansion. In the subsequent heat dissipation stage dominated by recrystallization after 20 ps, the amorphous molten region cools and shrinks accordingly. The protruding hillock slightly decreases in height owing to backflow and solidification of partial molten atoms. By the final steady state of 600 ps, an irradiation-induced hillock with a diameter of approximately 9.6 nm and a height of 2.1 nm is formed on the surface, leaving only scattered bubbles at the near-surface center of the ion track. This structure is referred to as a bell-shaped hillock,[23] as shown in Figure 2(c).

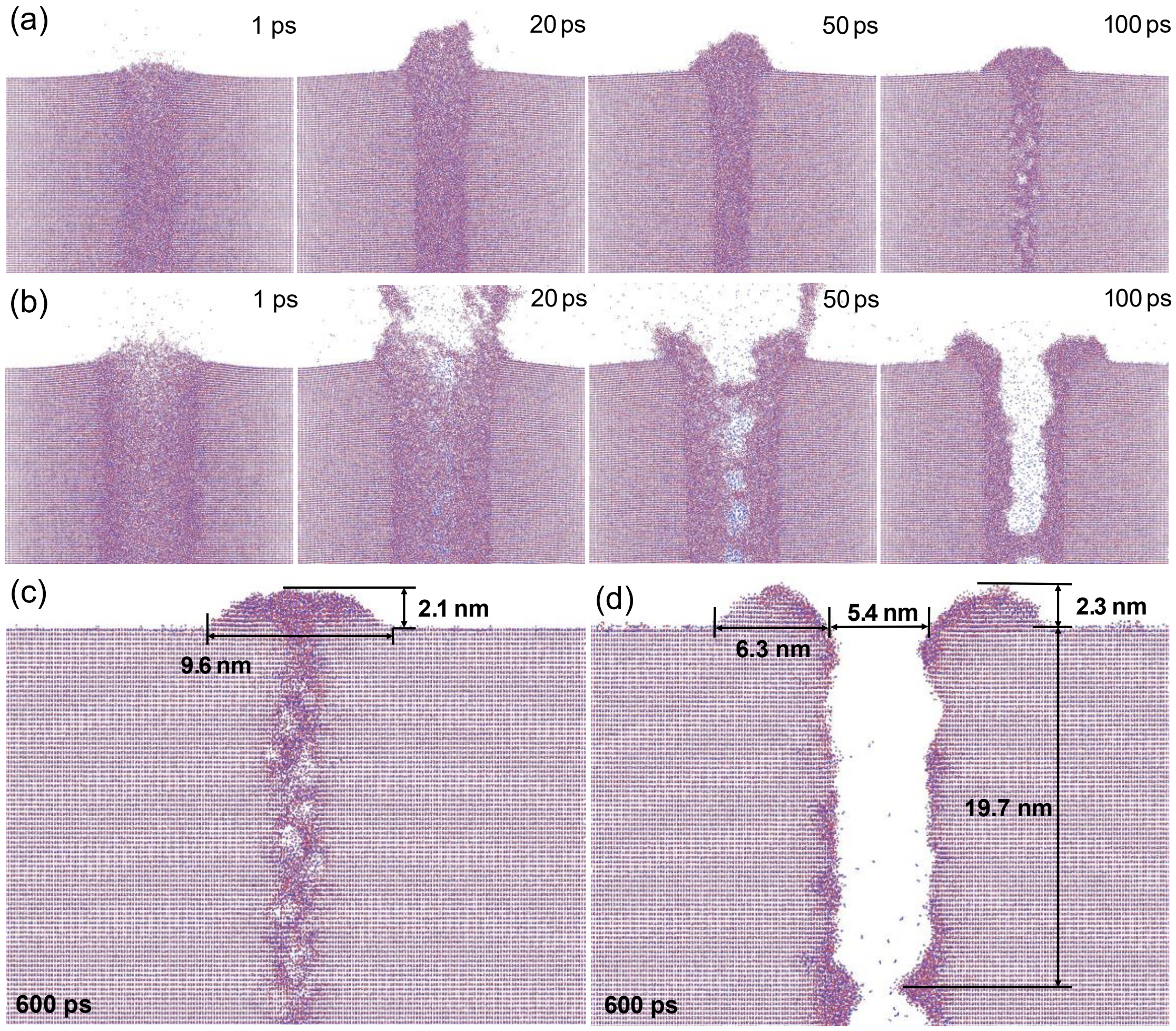

**FIG. 2.** Schematic evolution of surface hillock formation under (a) 430 MeV Kr and (b) 1171 MeV Ta irradiation. Final morphologies of surface hillocks under (c) 430 MeV Kr and (d) 1171 MeV Ta irradiation. Refer to supplementary materials section 2 to visualize the entire ion track formation process.

Compared with 430 MeV Kr irradiation, 1171 MeV Ta irradiation generates a much more intense thermal spike at the ion track center, yielding a significantly larger amorphous molten region and stronger pressure gradient. Atoms are strongly excited at 1 ps, accompanied by surface uplift and perturbation. Within the ensuing 20 ps, high pressure drives abundant molten atoms to sputter at high velocity, causing significant material loss in the central surface area. This rapid excavation collapses the original uplift into an incipient depression. As material is continuously stripped away, displaced molten atoms migrate laterally and accumulate along the pit edges, forming symmetric sidewall protrusions. Continuous sputtering facilitates the gradual excavation of a deep central cavity. At the final steady state of 600 ps, the surface morphology features a central continuous nanochannel flanked by symmetric protrusions, referred to as crater-rim hillocks,[24] as displayed in Figure 2(d). The central pit reaches a depth of 19.7 nm and a width of 5.4 nm, while the bilateral hillocks stand 2.3 nm high with an outer-edge distance of roughly 6.3 nm. Fundamentally, hillock formation is a pressure-driven response, as the ultrahigh thermal spike generated by SHI incidence creates a localized high-pressure environment, forcing internal molten atoms to sputter and extrude toward the free surface. Besides, the magnitude of deposited energy determined by $S_e$ values dictates distinct evolutionary pathways.

We have previously elucidated that SHI irradiation decomposes the wurtzite GaN structure into liquid Ga and $N_2$ molecules along the ion track.[22] Ga clusters and recrystallized wurtzite GaN mainly aggregate around these $N_2$ bubbles, effectively encapsulating $N_2$ molecules within the nanobubbles at the ion track center. As presented in Figure 3(a), apart from the atoms sputtered out of the system, a substantial amount of $N_2$ molecules remain at the ion track center after 1171 MeV Ta irradiation, whereas almost no $N_2$ molecules are retained following 430 MeV Kr irradiation. This contrast occurs because the weaker thermal spike under Kr irradiation induces minimal GaN melting and enables nearly complete lattice restoration, allowing any generated $N_2$ molecules in the near-surface region to easily escape and sputter away.

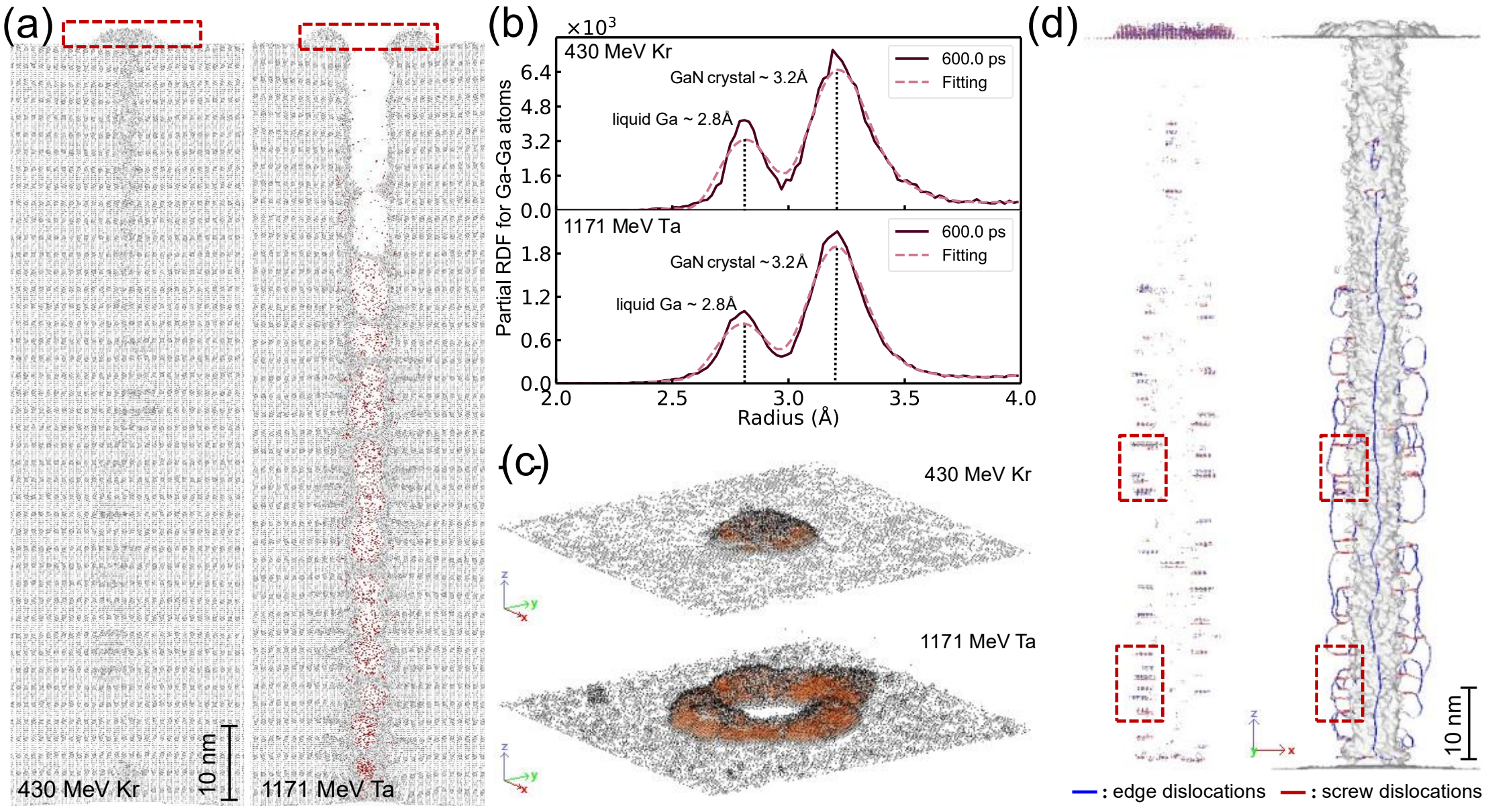

**FIG. 3.** Irradiation-induced microstructures in GaN under 430 MeV Kr and 1171 MeV Ta irradiation: (a) Spatial distribution of $N_2$ molecules (red spheres); (b) Partial RDF of Ga–Ga bonds in the hillock region indicated by the red box in (a); (c) Spatial distribution of zincblende GaN (red atoms) within the irradiation-induced hillocks. (d) Schematic illustration of zincblende nanodomains and dislocations induced by 1171 MeV Ta irradiation.

Owing to the rapid escape of $N_2$ molecules from the surface during intense sputtering, the extruded material that constitutes surface hillocks is inherently Ga-rich. Further atomic-level analysis corroborates this compositional characterization. As shown in Figure 3(b), the radial distribution function (RDF) of Ga–Ga bonds illustrate a distinct double-peak structure at approximately 2.8 Å and 3.2 Å under both irradiation conditions. The 2.8 Å peak aligns with the characteristic bond length of liquid Ga,[25,26] while the 3.2 Å peak matches the lattice constant of solid GaN.[27] This suggests that the irradiation-induced hillocks are entirely composed of liquid Ga clusters and recrystallized GaN.

In addition, the recrystallization process within these highly dynamic surface regions is not always perfect. Rapid recrystallization frequently induces the formation of zincblende GaN, a metastable phase with a face-centered cubic (FCC) configuration.[22] Structural characterization of the irradiation-induced hillocks reveals that a considerable amount of zincblende GaN is embedded within both the bell-shaped hillock induced by 430 MeV Kr irradiation and the crater-rim hillock induced by 1171 MeV Ta irradiation, with their distribution displayed in Figure 3(c). Apart from the surface hillocks, zincblende nanodomains also exist within the GaN bulk along the ion track. As exemplified by the 1171 MeV Ta irradiation in Figure 3(d), their distribution exhibits a strong spatial coincidence with the generated screw dislocations.[22]

Moreover, the synergistic effects of irradiation and temperature dominate the morphological evolution of surface hillocks, as presented in Figure 4. While the vertical heights of these hillocks generally stabilize within a range of 2 to 6 nm regardless of their specific morphological types as the temperature rises, their lateral dimensions expand markedly. Specifically, for the bell-shaped hillocks induced by 430 MeV Kr irradiation, the diameter increases monotonically from $9.6 \pm 0.2$ nm at 300 K to $17.7 \pm 0.2$ nm at 1800 K. The irradiation-induced zincblende nanodomains are predominantly concentrated on the periphery of bell-shaped hillocks. This distribution is consistent with the recrystallization process, where rapid cooling at the outer edge favors the formation of the metastable zincblende phase.[22] For the crater-rim hillocks induced by 1171 MeV Ta irradiation, the central-hole feature is preserved across all temperatures. Meanwhile, the diameter of the central hole shows a positive correlation with temperature, expanding from 5.4 ± 0.2 nm at 300 K to 13.4 ± 0.2 nm at 1800 K, with irradiation-induced zincblende nanodomains accumulating along the sidewalls. This temperature-dependent dimensional enlargement occurs because higher temperatures enhance the thermal spike generated by SHI irradiation. Consequently, the molten region expands and the subsequent cooling process decelerates. This extended thermal duration provides the highly fluid molten atoms with more time to migrate and accumulate at the surface, ultimately enlarging the hillock dimensions and aggravating the macroscopic irradiation damage.

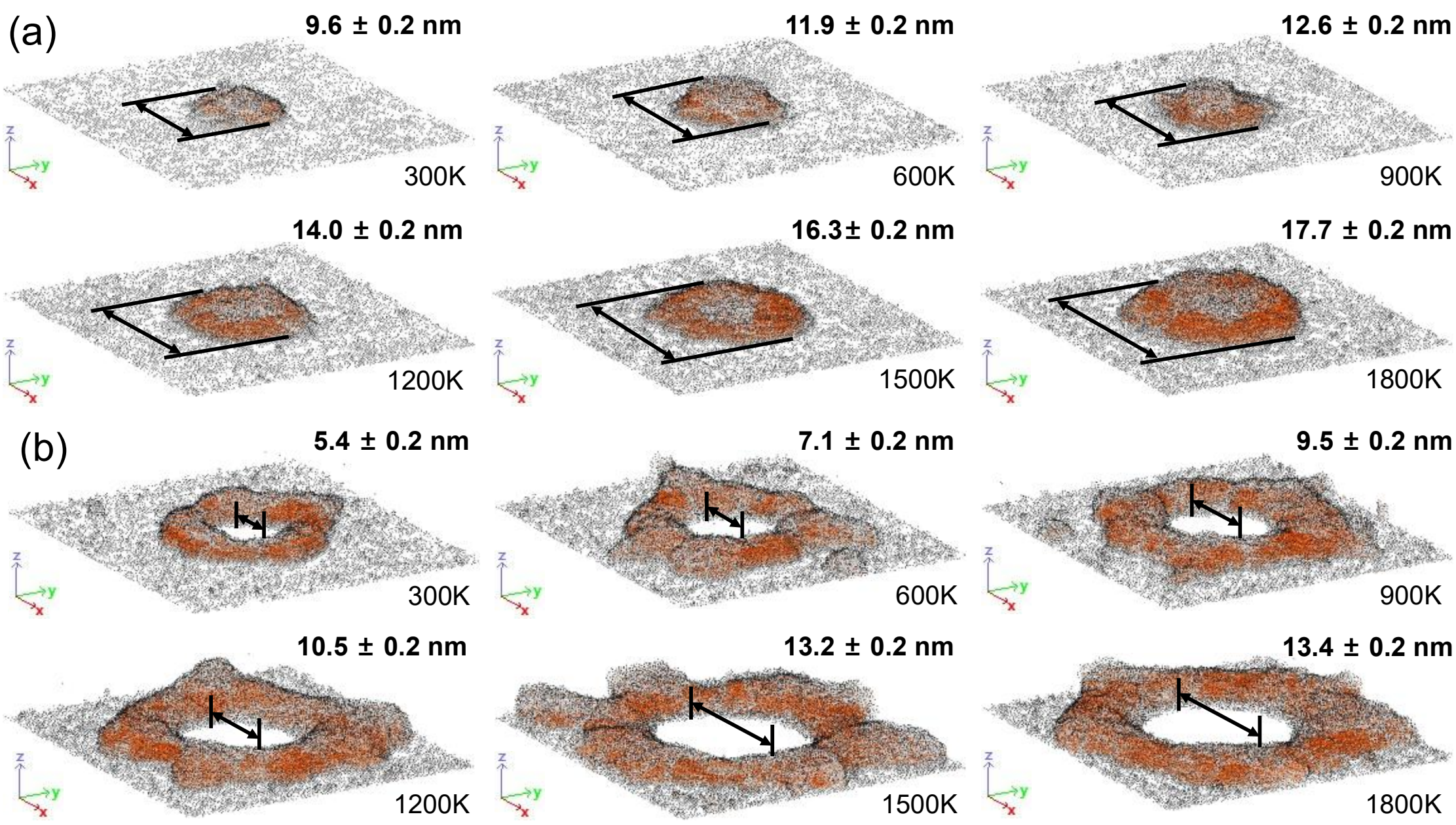

**FIG. 4.** Hillock morphologies under (a) 430 MeV Kr and (b) 1171 MeV Ta irradiation at different temperatures. The red regions represent zincblende GaN.

In parallel, this synergistic thermal effect drives the morphological evolution of the internal ion tracks extending deep into the bulk, as displayed in Figure 5 for GaN under 430 MeV Kr and 1171 MeV Ta irradiation across a temperature range from 300 K to 1800 K. For 430 MeV Kr irradiation, the size and distribution depth of scattered bubbles at the ion track center continuously increase as temperature rises and even coalesce into a continuous bubble channel. At 1800 K, this bubble channel extends to a length of approximately 25.3 nm near the surface. For 1171 MeV Ta irradiation, both the opening diameter and the depth of the pit increase gradually with temperature. At 1200 K, the pit almost penetrates the entire track depth and appears as a continuous nanochannel. Beyond this temperature threshold, the thermal effect mainly manifests as a slight expansion in the width of the continuous nanochannel.

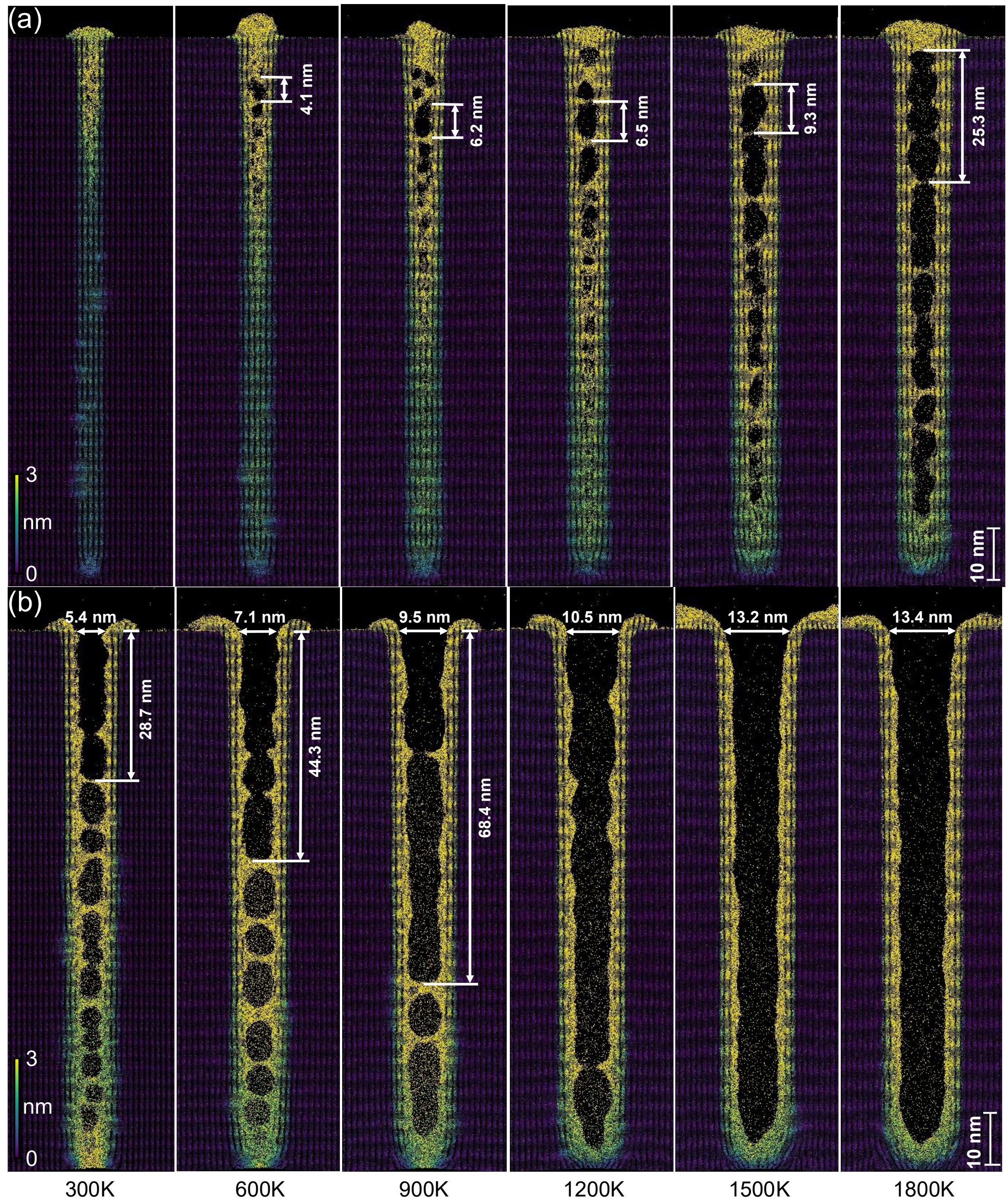

**FIG. 5.** Ion track morphologies under (a) 430 MeV Kr and (b) 1171 MeV Ta irradiation at different temperatures. The color scale represents the atomic displacement magnitude relative to the original lattice positions.

The underlying physical mechanism for this high temperature channel penetration can be attributed to the rheological properties of the melt phase. Xu *et al.* reported that the viscosity and surface tension of the transient molten phase are critical factors governing the formation of nanopores.[28] Specifically, lower viscosity and surface tension facilitate the outflow of molten material under the extreme pressure induced by the thermal spike, thereby promoting the formation of penetrating nanopores.[29,30] In the present case, the intense thermal spike generated by SHI irradiation decomposes GaN into liquid Ga and $N_2$ molecules. As $N_2$ molecules escape, the transient molten region formed near the surface is essentially a low-surface-tension system dominated by liquid Ga.[31,32] Furthermore, liquid Ga exhibits excellent fluidity due to its intrinsically low viscosity, which further decreases with increasing temperature.[33,34] This enhanced fluidity at elevated temperatures perfectly accounts for the formation and widening of penetrating nanochannels at 1200 K and above. Further temperature-dependent results including dislocations and zincblende nanodomains are provided in Section 3 of the supplementary materials.

This study reveals the atomistic formation mechanisms of bell-shaped and crater-rim hillocks under varying SHI irradiation conditions, alongside an in-depth analysis of their microstructural composition and structural characteristics. Furthermore, it elucidates the dynamic morphological evolution of ion tracks and surface hillocks driven by the synergistic effect of irradiation and transient high temperatures. These findings bridge the atomistic damage mechanisms with macroscopically observable surface degradation, providing a solid predictive basis for evaluating the resilience and reliability of GaN-based devices in extreme irradiation environments.

**SUPPLEMENTARY MATERIAL**

See the supplementary materials for detailed descriptions of the simulation setup and additional synergistic effects results.

## ACKNOWLEDGMENT

Financial support for this work was provided by the National Natural Science Foundation of China (NSFC) (No. 12405298), China Postdoctoral Science Foundation funded project (No. 2024M762610), and Innovation Scientific Program of CNNC.

## AUTHOR DECLARATIONS

### Conflict of Interest

The authors have no conflicts to disclose.

### Author Contributions

**Jiayu Liang:** Conceptualization (lead); Data curation (lead); Formal analysis (lead); Investigation (lead); Methodology (lead); Validation (lead); Visualization (lead); Writing – original draft (lead); Writing – review & editing (lead). **Shaowei He:** Investigation (supporting); Methodology (supporting). **Wenlong Liao:** Software (lead). **Tan Shi:** Data curation (supporting). **Hang Zang:** Visualization (supporting); Formal analysis (supporting). **Yonghong Li:** Investigation (supporting). **Wenbo Liu:** Formal analysis (supporting). **Xiaojun Fu:** Methodology (supporting). **Chuanjian Yao:** Methodology (supporting); Supervision (lead). **Huan He:** Conceptualization (supporting); Data curation (supporting); Funding acquisition (lead); Resources (lead); Writing – review & editing (supporting). **Jianan Wei:** Funding acquisition (supporting); Resources (supporting); Writing – review & editing (supporting). **Chaohui He:** Funding acquisition (supporting); Resources (supporting); Writing – review & editing (supporting).

## DATA AVAILABILITY

The data that supports the findings of this study are available from the corresponding authors upon reasonable request.